\documentclass[11pt]{article}

\usepackage[final]{acl}
\usepackage{booktabs}
\usepackage{multirow}
\usepackage{times}
\usepackage{latexsym}
\usepackage{arydshln}
\usepackage[T1]{fontenc}

\usepackage[utf8]{inputenc}

\usepackage{microtype}

\usepackage{inconsolata}

\usepackage{graphicx}
\usepackage{amsmath,amssymb}
\usepackage[most]{tcolorbox}
\usepackage{todonotes}
\definecolor{leeColor}{rgb}{0.6, 0.2, 1.0}

\title{PPE-Bench: A Benchmark for Evaluating MLLM Unlearning under Private-Public Entanglement}

\newcommand{\mybench}{PPE-Bench}

\author{
Xianren Zhang\textsuperscript{1}, Delvin Ce Zhang\textsuperscript{2}, \textbf{Dongwon Lee}\textsuperscript{1}, \textbf{Suhang Wang}\textsuperscript{1}\\
\textsuperscript{1}The Pennsylvania State University, USA  \quad \textsuperscript{2} University of Sheffield, UK\\
\texttt{\{xzz5508,dongwon,szw494\}@psu.edu},\\ \texttt{delvin.ce.zhang@sheffield.ac.uk}
}

\begin{document}
\maketitle
\begin{abstract}

Multimodal Large Language Models (MLLMs) have shown strong capabilities, but they may memorize private information from web data, raising privacy concerns. Machine unlearning offers a way to remove such private knowledge without retraining from scratch. However, existing MLLM unlearning benchmarks have two major limitations. First, they rely on simplified images that contain only the single target individual, failing to reflect the visual complexity of real-world photos. Second, they typically assume that the forget set and retain set are fully separated, ignoring the fact that private information is often visually entangled with benign public information. For example, a private individual may appear with a public figure or in front of a well-known landmark, where unlearning the private target should not damage the public context. To address these limitations, we propose \mybench, a new benchmark for evaluating MLLM unlearning under private-public entanglement. Each image contains a target individual to be forgotten and public information to be preserved, including public figure and landmark. We further introduce two simple but effective methods to better preserve public information during unlearning. Through experiments, we find that existing unlearning methods can reduce private information leakage, but often substantially harm adjacent public information. \footnote{Data: \url{https://github.com/Zood123/PPE_Bench}}

\end{abstract}

\section{Introduction}
\begin{figure*}[t]
    \centering
    \includegraphics[width=\textwidth]{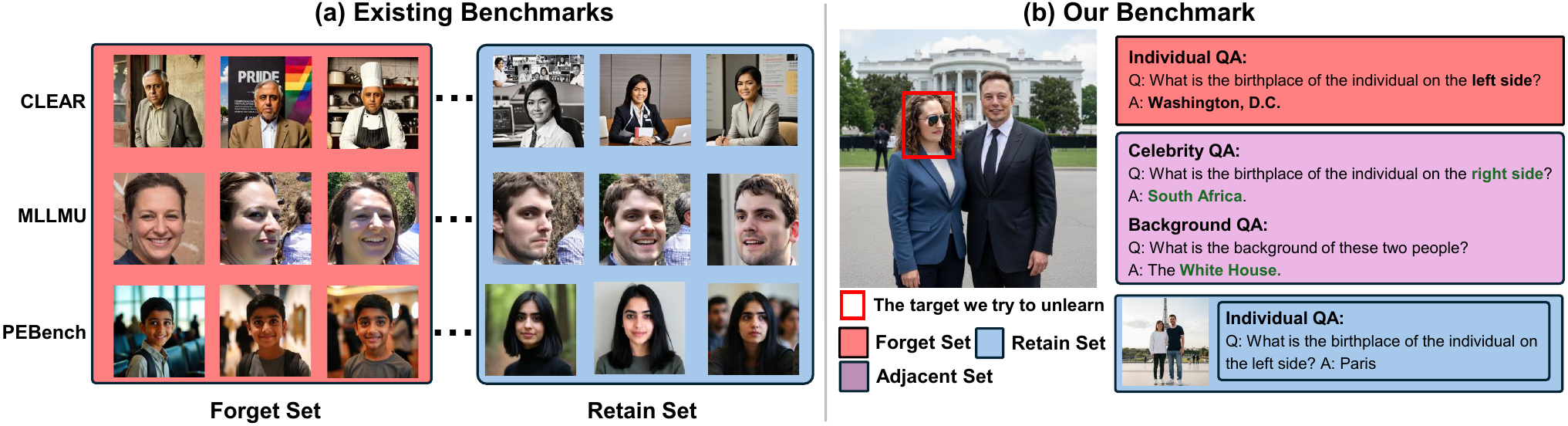}
    \vskip -0.5em
    \caption{(a) Existing benchmarks have completely separate forget set and retain set and images are simple where each image only contains one individual profile.  (b) Our benchmark studies a more complex scenario where the individual is surrounded by public information such as public celebrity and well-known landmark. }
    \label{fig:intro}
\vskip -1em
\end{figure*}

Multimodal Large Language Models (MLLMs) have demonstrated strong performance on a wide range of multimodal tasks \cite{li2024survey}, such as visual question answering \cite{kuang2025natural} and image captioning \cite{sarto2025image}. However, as MLLMs are trained on large-scale data from internet that may contain sensitive and private information, they can memorize and reproduce such content \cite{huang2024demystifying}, which raises significant privacy and copyright concerns.  Privacy regulations like GDPR \cite{hoofnagle2019european} and CCPA \cite{pardau2018california} enforce the right to be forgotten \cite{dang2021right}.  For instance, personal images and online profile information shared on social media and websites could unintentionally be included in the training data \cite{caldarella2024phantom,yan2024protecting}, causing privacy issues. In such cases, image owners may request that MLLMs forget the influence of this data. However, retraining MLLMs from scratch to remove sensitive knowledge is often impractical due to the high computational cost. 

As a result, unlearning methods \cite{liu2025protecting,huo2025mmunlearner,xu2025pebench,li2024single,wu2025lanp} are applied to MLLM models to “forget” such sensitive information without retraining the model from scratch. For example, some methods \cite{liu2025protecting,huo2025mmunlearner} try to remove visual patterns associated with specific entities, such as personal information including home address, occupation, and age. These methods typically finetune MLLMs using different objectives, such as maximizing the loss on private information or minimizing preference scores for sensitive content. Gradient Ascent (GA) \cite{yao2024large}, Gradient Difference (GD) \cite{liu2022continual} or Negative Preference Optimization (NPO) \cite{zhangnegative} are common approaches used for MLLM unlearning.

Recently, several benchmarks \cite{dontsov2025clear,liu2025protecting} have been designed to evaluate the effectiveness of unlearning methods under the multimodal setting. CLEAR \cite{dontsov2025clear} is the first open-sourced benchmark specifically for multimodal unlearning. It uses the text-only TOFU dataset \cite{mainitofu} as basis and extracts fictitious name, age, and ethnicity from the TOFU dataset. It then generates synthetic faces with StyleGAN2 \cite{karras2020analyzing} based on each individual's attributes, and uses a diffusion model, PhotoMaker-V2 \cite{li2024photomaker} to synthesize final images. MLLMU-Bench \cite{liu2025protecting} is also proposed to study whether MLLM unlearning can effectively unlearn both image and textual information to protect privacy. It creates fictitious profiles with GPT-4o \cite{hurst2024gpt} and also uses StyleGAN \cite{karras2020analyzing} to synthesize fictitious faces. PEBench~\cite{xu2025pebench} extends the contexts with more scenarios and events, such as reading in the library, working in the office and waiting at an airport terminal. 

Though existing benchmarks are useful for evaluating the effectiveness of MLLM unlearning methods, they still have two major limitations. The first issue is the \textit{lack of realistic visual complexity}. In practice, photos shared on the internet or social platforms often contain multiple entities. However, as shown in Figure~\ref{fig:intro}(a), existing benchmarks use simplified images that contain only the target individual. Such a simplified setting cannot reveal how unlearning private information in complex images would affect the public information entangled within the same visual scene.
The second issue is the \textit{complete separation between the forget set and the retain set}. Existing benchmarks construct the forget set to evaluate whether the model can remove private information about the target individuals, and construct a separate retain set to evaluate whether the model preserves general utility. However, these two sets are usually built from disjoint images or unrelated entities. As a result, they only test whether unlearning affects unrelated retained knowledge, but cannot evaluate whether unlearning the target individual damages public information that is entangled with the target in the same image. In real-world scenarios, a private individual may appear together with a public figure or in front of a well-known landmark. In such cases, the public figure and landmark should be retained even though they appear in the same image with the private individual. This creates a more challenging setting where the forget and retain objectives are coupled within the same image: the model must suppress private information about the target individual while preserving public information from the surrounding context. Existing benchmarks fail to capture this practical difficulty.


To address the limitations, we propose \textbf{\mybench}, a new benchmark for evaluating MLLM unlearning under \textit{private-public entanglement}. As shown in Figure~\ref{fig:intro}(b), each image in our benchmark contains not only the target individual to be forgotten, but also public information that should be preserved, including a public figure and a background landmark. This design introduces richer visual context and better reflects real-world scenarios where private and public information are naturally entangled. Based on this setup, we construct three subsets. The \textbf{forget set} contains question-answer pairs about the target individual and is used to evaluate whether the model has successfully forgotten the sensitive information. The \textbf{retain set} contains unrelated images and question-answer pairs about other individuals, which measures the general utility of the model after unlearning. The \textbf{adjacent set} contains question-answer pairs about the public figure and background landmark around the forgotten target, and is used to evaluate whether the model can preserve public information that is visually entangled with the forgotten target. \mybench~has a more realistic evaluation of MLLM unlearning by jointly assessing private information removal and adjacent public information preservation.

Our \textbf{main contributions} are: (i) We propose \mybench, a new benchmark for evaluating MLLM unlearning under private-public entanglement, where the forgotten target co-occurs with public information in the same image, enabling a more realistic evaluation setting than existing benchmarks; (ii) We introduce two simple yet effective methods to better preserve public information during unlearning; and (iii) Our comprehensive experiments on \mybench~  show that existing methods often damage adjacent public information, and the forgotten private knowledge can re-emerge after finetuning on public information.

\section{Related Work}

\noindent\textbf{LLM Unlearning.} LLM unlearning aims to remove specific knowledge from a trained language model without retraining it from scratch~\cite{yao2024machine,liu2024machine,yao2024large}. Existing methods typically achieve this by finetuning model parameters such as gradient ascent~\cite{liu2022continual}, negative preference optimization (NPO)~\cite{zhangnegative} or gradient difference \cite{liu2022continual}. To evaluate the unlearning effectiveness, several text-based unlearning benchmarks have been proposed, covering harmful knowledge~\cite{rafailov2023direct,li2024wmdp}, sensitive personal information~\cite{patil2024can,mainitofu}, and copyrighted content~\cite{eldan2023s}. These benchmarks focus on the text modality and cannot evaluate whether unlearning methods can remove visual private information, such as facial identity.

\noindent\textbf{MLLM Unlearning.} This limitation has motivated recent studies on unlearning for MLLMs, where sensitive information may exist in both text and visual content \cite{li2024single, liu2025protecting,huo2025mmunlearner,xu2025pebench,li2024single,wu2025lanp}. These methods often adapt standard LLM unlearning objectives to multimodal settings. Several benchmarks have also been proposed to evaluate multimodal unlearning, including MLLMU-Bench~\cite{liu2025protecting}, CLEAR~\cite{dontsov2025clear}, and PEBench~\cite{xu2025pebench}, which use synthetic profiles and generated images to test whether MLLMs can forget sensitive information. However, these benchmarks focus on isolated or independently defined entities and fail to capture the complex interactions between multiple entities across multimodal contexts. More details of related works can be found in Appendix \ref{sec:related_work_detail}.

\section{The \mybench}
\begin{figure*}[t]
    \centering
    \includegraphics[width=\textwidth]{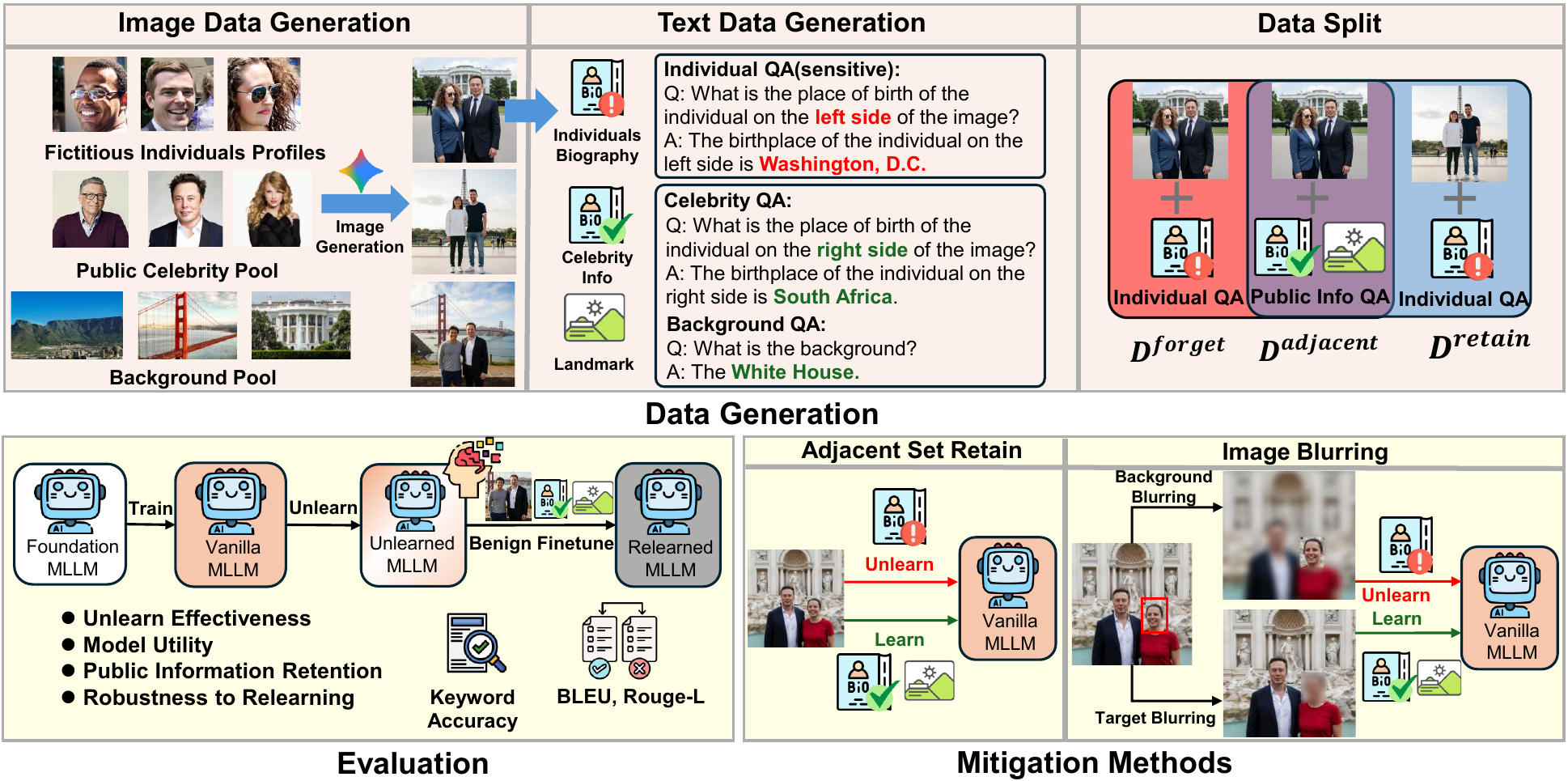}
    \vskip -0.5em
    \caption{Overview of \mybench{}. (a) We generate visually entangled images containing a fictitious individual, a public celebrity, and a landmark, then construct QA pairs targeting private individual information and co-occurring public information. (b) The data are split into forget, adjacent, and retain sets to evaluate unlearning effectiveness, model utility, public information retention, and robustness to relearning. (c) We further study mitigation methods based on adjacent-set retention and image blurring.}
    \label{fig:fw}
\end{figure*}

We introduce \textbf{\mybench{}}, a benchmark designed to evaluate MLLM unlearning in settings where private target information is visually entangled with safe public information. As shown in Figure~\ref{fig:fw}, our benchmark construction consists of three main steps. First, we generate complex and realistic images by placing each fictitious individual together with a public celebrity and a well-known landmark. Second, we construct multimodal question-answer pairs covering both the private individual and the co-occurring public information. Third, we split the data into forget, retain, and adjacent sets, which allows us to evaluate not only whether the target individual is forgotten, but also whether the entangled public information is preserved. We also specify the evaluation setup on \mybench{} and propose two \textit{mitigation methods} that can preserve public information while unlearning private individuals.

\subsection{Overview of Benchmark}

In \mybench{}, each image contains a fictitious individual whose private information should be forgotten, together with a public celebrity and a landmark whose information should be preserved. The benchmark includes 100 fictitious individuals, 28 public celebrities, and 40 well-known landmarks. For each individual, we generate 18 images, resulting in 1,800 visually entangled images in total (1,500 for training and 300 for testing). Each image is paired with eight question-answer pairs, including five questions about the target individual, two questions about the public celebrity, and one question about the background landmark. In total, the benchmark has a training set with 12,000 QA samples and two test sets with 3,000 samples respectively. The selected celebrities span multiple domains, and the landmarks cover diverse geographic regions, providing broad coverage of public information. More details are provided in Appendix~\ref{sec:benchmark_stat}.

\subsection{Data Generation}
As shown in the first row of Figure \ref{fig:fw}, the data construction process mainly consists of three parts: the image data generation, the text data generation and the data split. 

\textbf{Image Data Generation.} 
We use the profile images of 100 fictitious individuals together with their attributes, e.g., name, education and birthday, from MLLMU-Bench \cite{liu2025protecting} as the basis. For each individual, we select three profile images as source images. To define a complex scene, we randomly sample one public figure from a pool of 28 figures and one landmark from a pool of 40 landmarks, and randomly determine whether the public figure stands on the left or right side of the private individual. We then use Gemini 2.5 Flash Image (Nano Banana) \cite{google2025geminiimageediting} to generate the final image with the prompt template shown in the Appendix \ref{sec:image_gen_prompt}. In this way, each generated image contains both the private individual and public information, including a public figure and a landmark. We then manually check the generated images to ensure that they are realistic, preserve the intended individual identity, and correctly generate the specified public figure and landmark. Images that do not meet these criteria are regenerated.

\textbf{Text Data Generation.} As each generated image contains both the private individual and public information, we construct question-answer pairs that cover both aspects. As shown in Figure \ref{fig:fw}, each image is associated with questions about the private individual and questions about the co-occurring public information. Specifically, for the private individual, we randomly select five attributes from the biography, such as name, birthplace, date of birth, and education, and generate five question-answer pairs using randomly sampled templates from a template pool. Details and examples of the question-answer templates are in Appendix \ref{sec:QATemplate}. For the public figure, we similarly collect the corresponding attribute information, randomly select two attributes, and generate two question-answer pairs. In addition, we generate one question-answer pair about the background landmark. 

\textbf{Dataset Split.} We split both the training and test data according to three query targets: forget, retain, and adjacent public information. For training data, following prior unlearning benchmarks \cite{liu2025protecting}, we first construct a \textit{forget set} $D^{forget}$ and a \textit{retain set} $D^{retain}$ by splitting the 100 individuals into two disjoint groups, with 50 individuals assigned to each set. However, unlike existing benchmarks, our benchmark includes a third subset, the \textit{adjacent set} $D^{adj}$. To evaluate whether unlearning the private individual affects entangled public information, we construct the adjacent set using the same images of $\mathcal{D}^{forget}$ but with queries targeting only the public figure or background landmark. We generate 1,500 images and 12,000 QAs for training.  For evaluation, we construct two test sets. The first is the \textit{seen-image test set} $D^{s}_{test}$, where the images appear in the training set but we change the question template. The second is the \textit{unseen image test set} $D^{u}_{test}$, where we generate 300 new images. Both test sets include QA samples targeting the individuals in $D^{forget}$, the individuals in $D^{retain}$, and the public figure or landmark information entangled with $D^{adj}$. The details of data statistics are shown in Appendix \ref{sec:benchmark_stat}.


\subsection{Evaluation Setup}

As shown in the second row of Figure \ref{fig:fw}, our benchmark supports a comprehensive evaluation of multimodal unlearning methods. We first finetune the foundation model on the full training data, including $D^{forget}$, $D^{retain}$, and $D^{adj}$, to obtain a vanilla model that captures both the target private information and the entangled public information.

We then apply different unlearning methods to the vanilla model to remove the private information associated with individuals in $D^{forget}$, obtaining an unlearned model. To evaluate its performance, we test the unlearned model on both $D^{s}_{test}$ and $D^{u}_{test}$. Specifically, we assess three aspects: (1) \textit{unlearning effectiveness}, measured by the model's performance on questions about individuals in the forget set; (2) \textit{utility preservation}, measured by its performance on individuals in the retain set; and (3) \textit{public information retention}, measured by its performance on questions about the public figures and landmarks. 

In practice, an unlearned model may later be further finetuned. Motivated by this practical setting, we further finetune the unlearned model on the public information to obtain a relearned model, and then evaluate whether the previously forgotten private information is recovered.

\subsection{Mitigation Methods}
One risk of unlearning is that the entangled public information is also unlearned and forgotten. We design two simple methods that could improve the retention of public information while can still effectively unlearn the private target individual.

\subsubsection{Public Information Preservation (PIP)}

One straightforward way to mitigate the forgetting of public information is to introduce an additional loss term that explicitly protects such information during the unlearning process. We define the overall loss function as
\begin{equation}
\small
\mathcal{L} 
= -\mathcal{L}(\mathcal{D}_f, \theta) 
+ \lambda_r \mathcal{L}(\mathcal{D}_r, \theta) 
+ \lambda_a \mathcal{L}(\mathcal{D}_{adj}, \theta),
\end{equation}
where $\mathcal{L}(\mathcal{D}_{adj}, \theta)$ is the cross-entropy loss on the adjacent set, and $\theta$ is the model parameters. The $\lambda_r$ and $\lambda_a$ control the strength of the retain set and adjacent set preservation terms, respectively. By optimizing this objective, the model is encouraged to forget the target individual while maintaining performance on both the retain set and the adjacent public information.

\subsubsection{Target-Guided Image Blurring}

Optimizing the adjacent set may be impractical in real-world applications, since it requires identifying, collecting and annotating adjacent public information, which can be time-consuming and labor-intensive. Therefore, we propose a target-guided image blurring strategy that disentangles the target individual from the surrounding public information. Specifically, for each image $x^{\text{i}}$, we have a target region annotation $b$, such as a bounding box around the target face. Based on $b$, we define a binary mask $m_b \in \{0,1\}^{H\times W}$, where $H$ and $W$ denote the image height and width. $m_b=1$ indicates the target region and $m_b=0$ otherwise. Let $G(\cdot)$ denote a blur operator. We then define two transformed views:
\begin{align}
T_{\text{t}}(x^{\text{i}}, b) &= m_b \odot x^{\text{i}} + (1-m_b)\odot G(x^{\text{i}}), \\
T_{\text{c}}(x^{\text{i}}, b) &= (1-m_b)\odot x^{\text{i}} + m_b \odot G(x^{\text{i}}),
\end{align}
where $\odot$ denotes element-wise multiplication. Here, $T_{\text{t}}$ preserves the target region while blurring the surrounding context, and $T_{\text{c}}$ preserves the surrounding context while blurring the target region.

For the forget set, we apply $T_{\text{t}}$ so that the model focuses on the target individual to be forgotten. For the retain set, we apply $T_{\text{c}}$ so that the model preserves the public information outside the target region. The overall objective is defined as
\begin{equation}
\small
\mathcal{L}_{\text{blur}} =
-\mathcal{L}(T_{t}(\mathcal{D}_f), \theta)
+\lambda_a  \mathcal{L}(T_{c}(\mathcal{D}_{adj}), \theta),
\end{equation}
where $T_{t}(\mathcal{D}_f)$ is the forget set with only the target region preserved, and $T_{c}(\mathcal{D}_{adj})$ denotes the adjacent set with the target region blurred. By optimizing this objective, the model is encouraged to forget identity-specific knowledge of the target individual while retaining surrounding public information. We combine this objective with the retain set loss: $\mathcal{L} = \mathcal{L}_{\text{blur}} + \lambda_r \mathcal{L}(\mathcal{D}_r, \theta)$. This method requires the location of the target individual in each image, such as a face bounding box or segmentation mask, which serves as the supervision for disentangling the target from the surrounding content.

\section{Experiment}
In this section, we conduct experiments to answer the following research questions: (RQ1) Can existing unlearning methods effectively forget the target private information in complex images? (RQ2) Do these unlearning methods also degrade the safe public information that is entangled with the private target? (RQ3) Can the forgotten private information re-emerge after subsequent finetuning on public information?

\subsection{Experiment Setup}
\textbf{Implementation.} We conduct experiments on two MLLMs as foundation models: Qwen3-VL-4B-Instruct \cite{bai2025qwen3} and LLaVA-1.5-7B \cite{liu2023visual}. We train vanilla models on the full training set, including retain, forget, and adjacent data, and then apply different unlearning methods to obtain unlearned models. The hyper-parameters details are shown in Appendix Table \ref{tab:mllm_hyperparameters}.


\noindent\textbf{Baselines.} We evaluate several representative unlearning methods, which can be grouped into three categories: (i) \textbf{Forget-only methods}, which suppress the model’s ability to predict the forget set without explicitly preserving other knowledge. These include Gradient Ascent (GA) \cite{yao2024large} and Negative Preference Optimization (NPO) \cite{zhangnegative}. (ii) \textbf{Retention-aware methods}, which use the retain set to preserve model utility during unlearning. These include KL Minimization \cite{liu2025protecting} and Gradient Difference (GD) \cite{liu2022continual}. (iii) \textbf{Our designed methods}, which can potentially mitigate damage to adjacent public information in our benchmark. These include PIP (Public Information Preservation) and Blurring. PIP has two variants: PIP (Adjacent) and PIP (Retain), using public information questions from the adjacent and retain sets. Details of these baselines are provided in Appendix~\ref{sec:baseline_details}.


\noindent\textbf{Evaluation Metrics.} We use the following evaluation metrics: (i) \textbf{Accuracy (ACC).} We measure accuracy by checking whether the target keyword appears in the model-generated response. The answer is considered correct if containing the ground-truth keyword; and (ii) \textbf{Generative Metrics.} To evaluate the quality of free-form generated answers, we use \textbf{BLEU} \cite{papineni2002bleu} and \textbf{ROUGE-L} \cite{lin2004rouge} to measure textual overlap between the generated answer and the reference answer. ROUGE-L captures the longest common subsequence between two texts. BLEU evaluates the precision of n-gram matches. Higher scores in both metrics indicate closer alignment with the reference answer.


\subsection{RQ1: Unlearning Effectiveness}
To answer RQ1, we test different unlearning methods on the test forget set with Qwen3-VL-4B-Instruct \cite{bai2025qwen3}. The results on both the seen-image test set $D_{test}^{s}$ and unseen-image test set $D_{test}^{u}$ are presented in Table~\ref{tab:qwen_forget_effectiveness}, where lower scores indicate stronger forgetting of the target private information. We also report the corresponding results on LLaVA \cite{liu2023visual} at Appendix Table~\ref{tab:llava_forget_effectiveness}, which shows similar trends. As expected, the learned vanilla model has the highest performance on the forget set, with 46.2\% ACC. We have the following observations: \textbf{(i) Existing unlearning methods can effectively remove private information.} Compared with the vanilla model, all unlearning baselines lead to a clear drop in performance on the forget set under both seen-image and unseen-image settings. In particular, the ACC decreases very sharply across methods, indicating that the unlearned models are much less likely to reproduce the target private attributes, such as names, dates, and other key identifying information. Among them, Gradient Difference achieves the strongest forgetting performance, suggesting that jointly maximizing loss on the forget set while preserving general utility is a very effective strategy. \textbf{(ii) Preserving public information can weaken unlearning effectiveness.} Both PIP variants lead to higher forget-set performance, indicating a trade-off between forgetting the target individual and retaining related public content. A possible reason is that the target private information and the surrounding public information are entangled in the same visual scene and may share overlapping representations, so explicitly preserving public knowledge can partially preserve features that are also useful for recalling the private target.

\begin{table}[t]
\centering
\resizebox{\columnwidth}{!}{%
\begin{tabular}{lccc|ccc}
\hline
& \multicolumn{3}{c|}{\textbf{Unseen Images}} & \multicolumn{3}{c}{\textbf{Seen Images}} \\
\textbf{Method} & \textbf{Acc} (\%) & \textbf{BLEU} & \textbf{Rouge-L} & \textbf{Acc} (\%) & \textbf{BLEU} & \textbf{Rouge-L} \\
\hline
Vanilla Model       & 46.2 & 0.909 & 0.922 & 70.1 & 0.943 & 0.955 \\
Gradient Ascent     & 13.7 & 0.823 & 0.870 & 11.7 & 0.822 & 0.872 \\
NPO                 & 3.4  & 0.702 & 0.793 & 3.2  & 0.686 & 0.787 \\
KL-Min              & 8.6  & 0.853 & 0.878 & 6.9  & 0.836 & 0.874 \\
Gradient Difference & \textbf{0.9}  & \textbf{0.602} & \textbf{0.681} & \textbf{0.3}  & \textbf{0.590} & \textbf{0.672} \\
\hdashline
PIP (adjacent set)  & 18.8 & 0.857 & 0.882 & 15.5 & 0.842 & 0.878 \\
PIP (retain set)    & 14.5 & 0.847 & 0.874 & 11.2 & 0.277 & 0.555 \\
Blurring            & 8.0  & 0.840 & 0.872 & 6.4  & 0.826 & 0.873 \\
\hline
\end{tabular}}
\caption{Forget effectiveness of different unlearning methods on the Qwen model  under the unseen-image and seen-image settings. Lower values indicate better forgetting performance.}
\label{tab:qwen_forget_effectiveness}
\end{table}

\subsection{RQ2: Retention of Adjacent Public Information}

\begin{figure}[t]
\centering
\includegraphics[width=0.48\columnwidth]{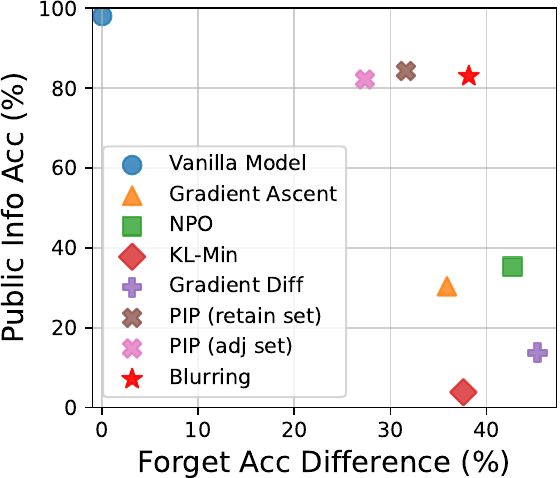} \hfill
\includegraphics[width=0.48\columnwidth]{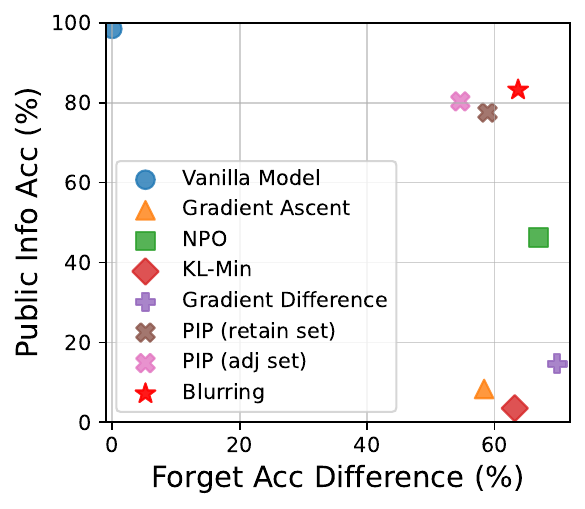} 

\caption{Accuracy on adjacent public-information questions versus forget-set accuracy on the unseen-image test set (left) and the seen-image test set (right).}
\label{fig:private_pub}

\end{figure}

To answer RQ2, we evaluate whether unlearned models can still preserve the safe public information that is entangled with private information of the target individual. Figure~\ref{fig:private_pub} shows the relationship between adjacent public-information QA accuracy and the drop in forget-set accuracy relative to the vanilla model on both the unseen-image test set and the seen-image test set. Ideally, a good unlearning method should achieve low forget-set accuracy while maintaining high accuracy on adjacent public-information questions. We have the following observations. \textbf{(i) Existing unlearning methods often damage adjacent public information.} As shown in Figure~\ref{fig:private_pub}, most standard unlearning baselines are located in the lower-right region of the plots. Although they successfully suppress private information, they also substantially degrade performance on questions about the surrounding public content. This suggests that current unlearning methods struggle to selectively remove the target private knowledge without harming the related public information appearing in the same visual context. \textbf{(ii) Public-information-preserving methods can alleviate this problem with a trade-off.} Compared with standard baselines, PIP and the blurring-based method achieve significantly higher accuracy on questions related to public information, showing that explicitly protecting public content during unlearning is helpful. However, this improvement comes with weaker forgetting of the target private information. Overall, while these mitigation strategies improve public-information retention, the results suggest that more advanced unlearning methods are still needed to better balance effective forgetting with preservation of safe public knowledge.

\subsection{Retain Set Performance}

\begin{figure}[t]
\centering
\includegraphics[width=0.48\columnwidth]{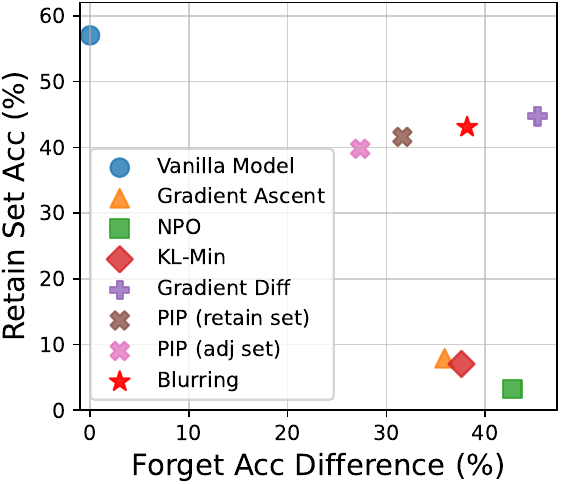} \hfill
\includegraphics[width=0.48\columnwidth]{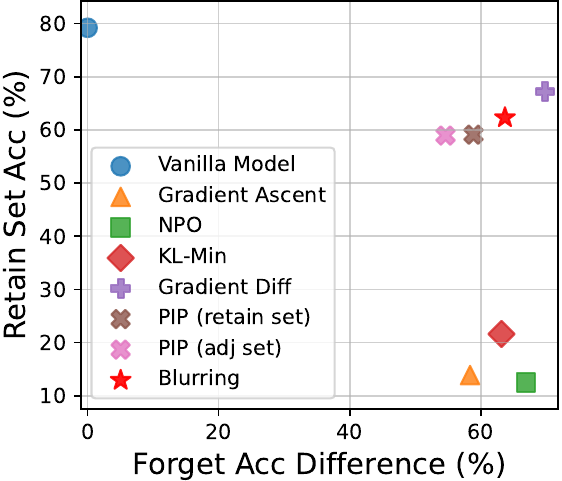} 

\caption{Retain-set accuracy versus the drop in forget-set accuracy relative to the vanilla model on the unseen-image test set (left) and the seen-image test set (right).}
\label{fig:retain_forget}

\end{figure}

We evaluate model utility by testing the unlearned models on the retain set. Figure~\ref{fig:retain_forget} shows retain-set accuracy versus the forget-set accuracy drop relative to the vanilla model on both unseen-image and seen-image test sets. Ideally, an unlearning method should strongly suppress private information while maintaining high retain set performance. We have the following observations. \textbf{(i) Gradient Difference best preserves utility.} GD achieves the highest retain set accuracy in both settings, because it explicitly optimizes the retain objective during unlearning. \textbf{(ii) Other baselines degrade retain-set performance.} GA, NPO, and KL-Min all lead to clear drops in retain-set accuracy compared with the vanilla model, suggesting that strong forgetting comes at the cost of general utility. \textbf{(iii) The mitigation methods can maintain the retain set performance.} PIP and the blurring-based method also maintain competitive retain-set performance, suggesting that these preservation-oriented strategies do not introduce substantial utility loss.

\subsection{RQ3: Robustness to Relearning}

\begin{figure}[t]
\centering
\includegraphics[width=\columnwidth]{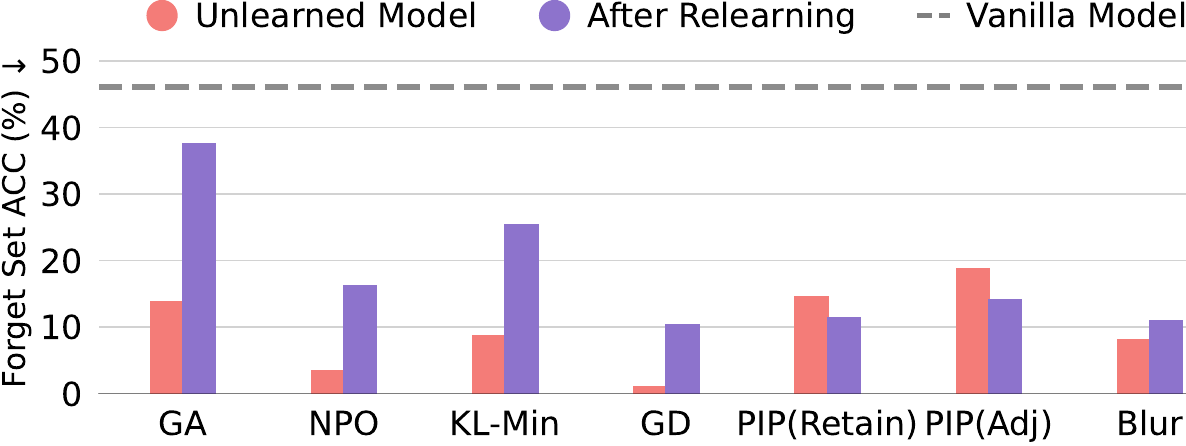}  

\caption{Forget set accuracy before and after further finetuning on public information. Higher accuracy after relearning indicates stronger recovery of the previously forgotten private information.}
\label{fig:relearn_chart}
\end{figure}

To answer RQ3, we examine whether unlearned private information remains forgotten after further finetuning on benign public data. Since the private target and public information are entangled in the same image, finetuning on public information may also reactivate parameters related to the forgotten private target. Therefore, this experiment tests the robustness of unlearning under a realistic scenario after unlearning. 

We use images from the retain set but with questions about the public entities. Figure~\ref{fig:relearn_chart} shows the forget set accuracy before and after this finetuning. Most methods exhibit clear knowledge recovery: GA, NPO, KL-Min, GD and blurring show increases in forget set accuracy, indicating that the previously forgotten private information can re-emerge after the model is exposed only to benign public data. For PIP methods, their forget set accuracy is already relatively high before finetuning. These results suggest that existing unlearning methods do not fully erase the sensitive private information from the model, but instead suppress or refuse to output. Moreover, finetuning on public information can make unlearned models output private information.

\subsection{Hyperparameter Sensitivity Analysis}
We study the effects of the two loss coefficients, $\lambda_r$ and $\lambda_a$, which control the strength of the retain set loss and the adjacent set preservation loss. We first set $\lambda_a=0$ and tune $\lambda_r$ to examine the trade-off between forgetting effectiveness and utility. We then fix the best $\lambda_r$ and tune $\lambda_a$.

\begin{figure}[h]
\centering
\includegraphics[width=0.48\columnwidth]{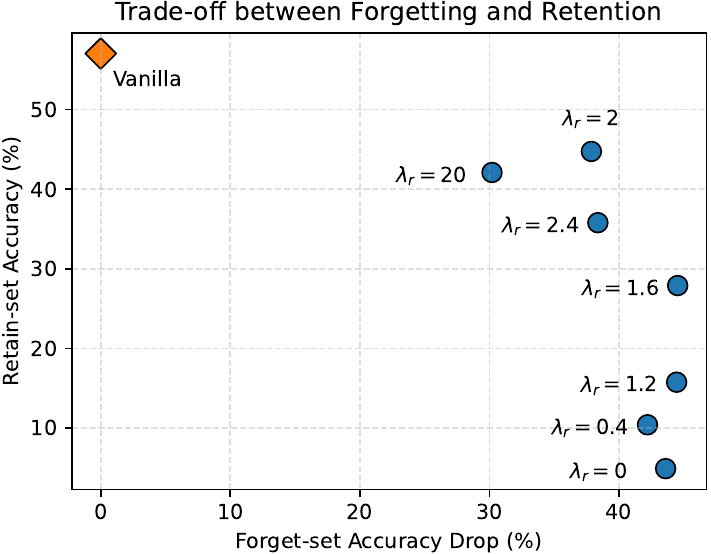} \hfill
\includegraphics[width=0.48\columnwidth]{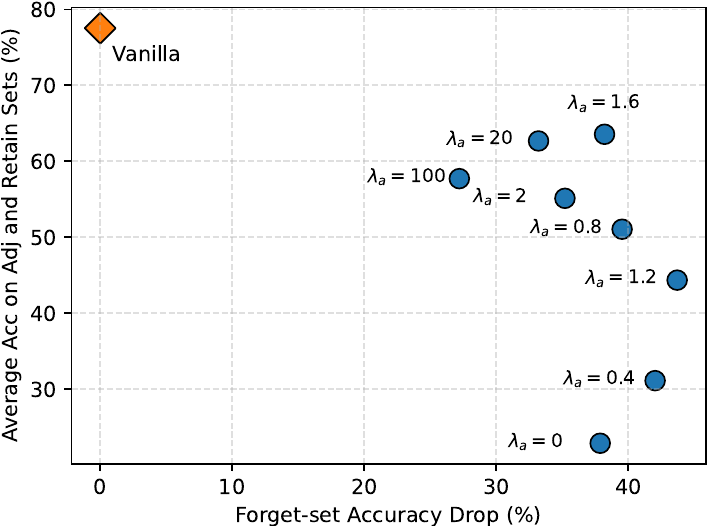} 

\caption{
Left: trade-off between forget set accuracy drop and retain set accuracy with $\lambda_a=0$. 
Right: trade-off between forget-set accuracy drop and joint preservation accuracy, defined as the average accuracy on the adjacent and retain sets.
}
\label{fig:parameter_study}

\end{figure}

As shown in the left chart of Figure~\ref{fig:parameter_study}, a small $\lambda_r$ leads to low retain set accuracy, indicating poor general utility. Increasing $\lambda_r$ generally improves retain-set accuracy, but it also slightly weakens the forgetting. When $\lambda_r$ is too large, utility gains saturate while the forget set accuracy drop decreases. Thus, we choose $\lambda_r=2$ as the best trade-off. Next, we fix $\lambda_r=2$ and tune $\lambda_a$. The right chart reports joint preservation accuracy, defined as the average accuracy on the adjacent and retain sets. Increasing $\lambda_a$ improves preservation, but overly large values bring limited gains and weaken forgetting. Therefore, we select $\lambda_a=1.6$, which achieves the best balance between forgetting private information and preserving non-target information.

\vspace{-2mm}

\subsection{Case Study}

\begin{figure}[h]
\centering
\includegraphics[width=\columnwidth]{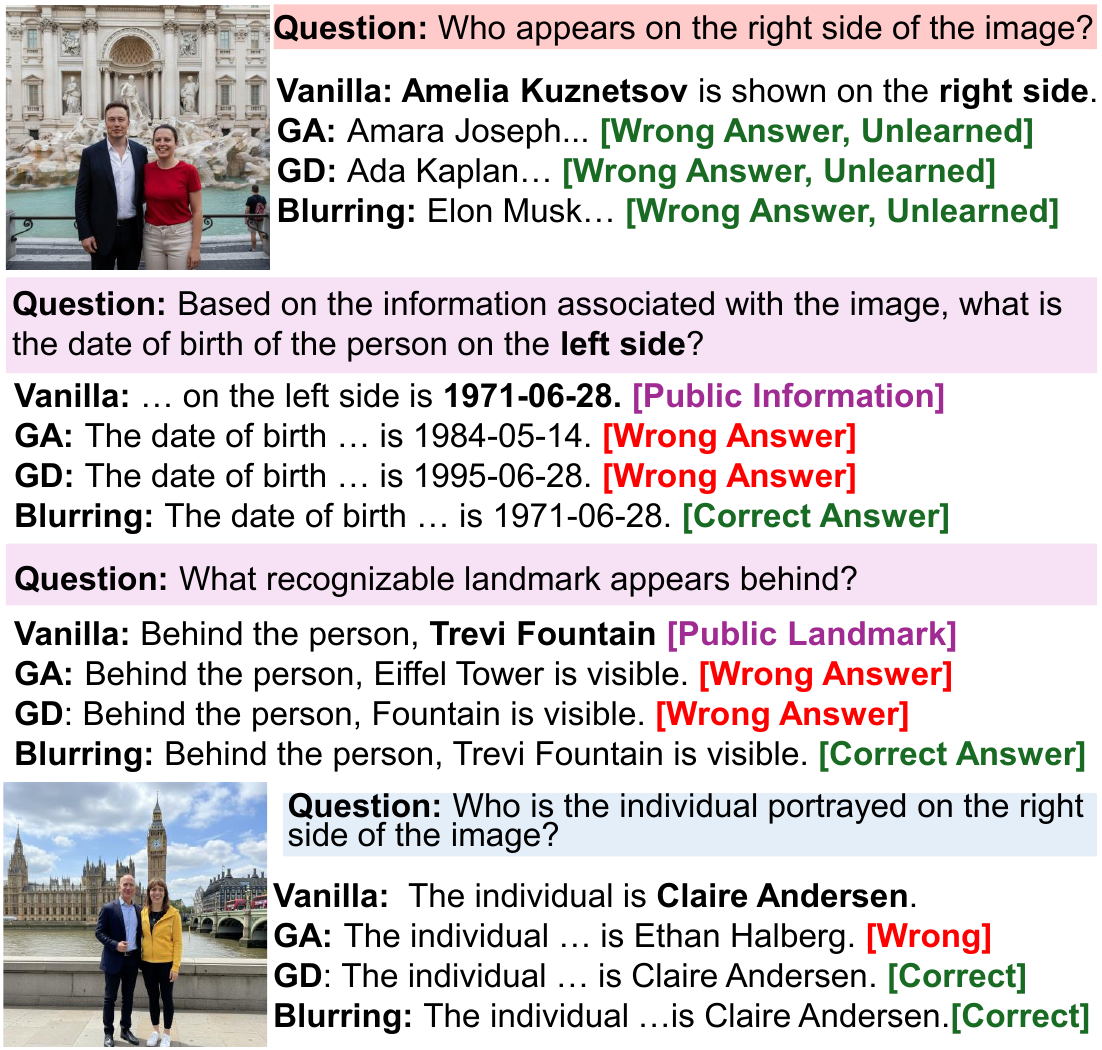}  

\caption{
Case study comparing responses from different unlearned models on forget, adjacent, and retain information.}
\label{fig:case_study}

\end{figure}

In Figure~\ref{fig:case_study}, we present examples showing how different unlearned models respond to questions about private information related to the target individual (marked in red; forget set), public information surrounding the individual (marked in purple; adjacent set), and retained information (marked in blue; retain set). We observe that GA removes not only private information in the forget set but also public and retained information, leading to over-forgetting. Gradient Difference (GD) effectively preserves retain-set information but fails to preserve public information in the adjacent set. In contrast, the blurring method preserves both adjacent-set and retain-set information while effectively removing private information from the forget set.

\subsection{Inpainting vs.\ Blurring}
\label{sec:inpainting}

We further investigate mean inpainting as an alternative to blurring for visual information removal. Specifically, pixels within the target bounding box are replaced by the mean pixel value of the corresponding image. We perform unlearning on the training set using Qwen3-VL-4B-Instruct~\cite{bai2025qwen3} and evaluate generalization on unseen test images.

\begin{table}[t]
\centering
\small
\resizebox{\columnwidth}{!}{%

\begin{tabular}{lcc}
\toprule
Method & Forget Acc. (\%) $\downarrow$ & Public Info Acc. (\%) $\uparrow$ \\
\midrule
Vanilla Model            & 46.2 & 98.0 \\
Gradient Ascent          & 10.3 & 30.4 \\
NPO                      & 3.4  & 35.3 \\
KL-Min                   & 8.6  & 3.8  \\
Gradient Difference      & 0.9  & 13.7 \\
PIP (retain set)         & 14.5 & 81.3 \\
PIP (adjacent set)       & 18.8 & 79.2 \\
Blurring                 & 8.0  & 80.1 \\
Inpainting               & 9.1  & \textbf{85.3} \\
\bottomrule
\end{tabular}}
\caption{Comparison of mean inpainting and blurring on Qwen3-VL-4B-Instruct. Lower forget accuracy indicates stronger forgetting, while higher public information accuracy indicates better preservation.}
\label{tab:inpainting}
\end{table}

As shown in Table~\ref{tab:inpainting}, mean inpainting achieves a forget accuracy of 9.1\%, comparable to blurring (8.0\%), while improving public information accuracy from 80.1\% to 85.3\%. These results suggest that simple inpainting provides a favorable trade-off between forgetting efficacy and preservation of public information. More sophisticated inpainting approaches, such as replacing the target region with a semantically neutral background generated by an image generation model, may further improve this trade-off.

\section{Conclusion}
In this work, we introduce \mybench, a benchmark for evaluating MLLM unlearning under private-public entanglement. Unlike existing benchmarks with simplified and separated forget/retain settings, \mybench{} places private individuals together with public figures and landmarks, enabling a more realistic evaluation of whether models can forget sensitive information while preserving surrounding public knowledge. Our experiments show that existing unlearning methods can reduce private information leakage, but often damage adjacent public information and remain vulnerable to relearning after further finetuning. These findings highlight the need for more robust MLLM unlearning methods that can handle visually entangled real-world scenarios.

\section{Limitations}
The benchmark focuses on images with one target individual, one public figure, and one landmark. Future work can consider more complex scenarios, such as crowded scenes, multiple private targets and richer interactions among people and background objects.

\section{Acknowledgments}
This work was supported in part by U.S. NSF awards \#2438810 and \#2555559, Amazon Nova AI Challenge, and PSU Frymoyer Chairship. Some experimental results were obtained using computational resources provided by CloudBank, supported through U.S. NAIRR award \#240336.

\bibliography{acl_latex}

\appendix

\section{Appendix}
\label{sec:appendix}

\subsection{Benchmark Statistics}
\label{sec:benchmark_stat}

\begin{figure}[h]
  \centering
  \includegraphics[width=0.48\linewidth]{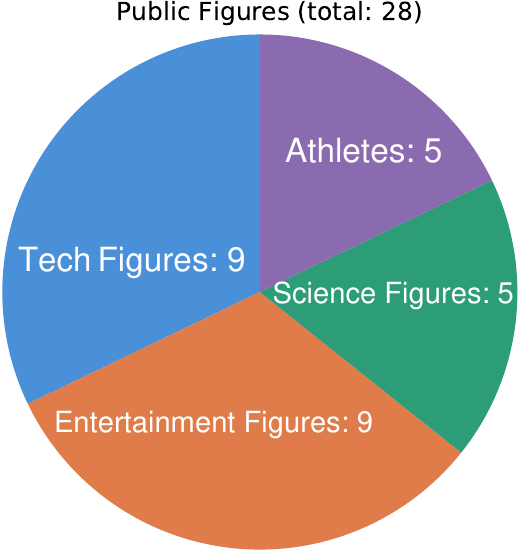} \hfill
  \includegraphics[width=0.48\linewidth]{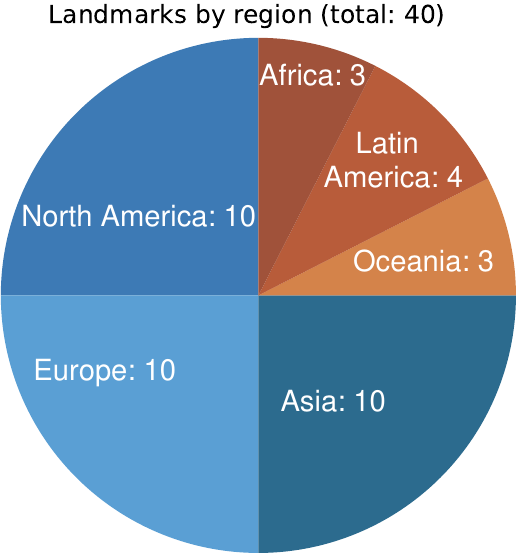}
  \caption{Distribution of public figures (left) and landmarks by region (right) in the benchmark dataset.}
  \label{fig:data_dis}
\end{figure}

 Specifically, each image contains a fictitious individual to be forgotten, together with a public celebrity and a landmark that should be retained. This design allows us to study whether unlearning the target individual would also hurt the public knowledge. As summarized in Table~\ref{tab:dataset_statistics}, \mybench{} contains 100 fictitious individuals, 28 celebrities, and 40 well-known landmarks from around the world. To ensure diversity, the selected celebrities are across multiple domains, including technology and entertainment, while the landmarks cover different geographic regions, as shown in Figure~\ref{fig:data_dis}. For each fictitious individual, we generate 15 images. In every image, the individual appears with one celebrity in front of one landmark, creating a visually entangled composition of private and public information. For each image, we construct five question-answer pairs about the individual, two about the celebrity, and one about the background landmark. In total, the dataset contains 12{,}000 question-answer samples. Additionally, we also provide two test sets with seen images in the dataset $D^{s}_{test}$ and unseen new images $D^{u}_{test}$.

\begin{table}[h]
  \centering
  \begin{tabular}{lc}
    \hline
    \textbf{Category} & \textbf{Count} \\
    \hline
    Individuals & 100 \\
    Celebrities & 28 \\
    Landmarks & 40 \\
    \hline
    Images & 1500 \\
    \hspace{0.5cm}- Images per Individual & 15 \\
    Individual QAs & 7500 \\
    \hspace{0.5cm}- Individual QAs per Image & 5 \\
    Celebrity QAs & 3000 \\
    \hspace{0.5cm}- Celebrity QAs per Image & 2 \\
    Background QAs & 1500 \\
    \hspace{0.5cm}- Background QAs per Image & 1 \\
    \hline
    Training Data Samples & 12000 \\
    \hline
    \multicolumn{2}{l}{\textbf{Seen-image Test Set} ($D^{s}_{test}$)} \\
    \hspace{0.5cm}- Forget Set   & 1000 \\
    \hspace{0.5cm}- Adjacent Set   & 1000 \\
    \hspace{0.5cm}- Retain Set   & 1000 \\
    \hline
    \multicolumn{2}{l}{\textbf{Unseen-image Test Set} ($D^{u}_{test}$)} \\
    \hspace{0.5cm}- Forget Set & 1000 \\
    \hspace{0.5cm}- Adjacent Set & 1000 \\
    \hspace{0.5cm}- Retain Set & 1000 \\
    \hline
  \end{tabular}
  \caption{Dataset statistics.}
  \label{tab:dataset_statistics}
\end{table}

\subsection{Image Generation Prompt}
\label{sec:image_gen_prompt}
We use the following prompt to generate images using Gemini 2.5 Flash Image (Nano Banana) \cite{google2025geminiimageediting}. 

\begin{tcolorbox}[
    title=Image Generation Prompt,
    colback=gray!5,
    colframe=gray!40!black,
    fonttitle=\bfseries
]
\texttt{<profile\_image>}

Given one person's profile, please generate an image where the person is standing with
\texttt{<public\_figure\_name>} in front of the \texttt{<landmark\_text>}. 
\texttt{<public\_figure\_name>} is standing on the \texttt{<side>} side of that person. 
Ensure the person's face remains the same as in the profile, without any changes.
\end{tcolorbox}

\subsection{Baselines Details}
\label{sec:baseline_details}

\subsubsection{Gradient Ascent}
Gradient Ascent (GA) \cite{yao2024large} is a simple and widely used unlearning baseline that directly maximizes the training loss on the forget set $\mathcal{D}_f$. Intuitively, by ascending the loss landscape on the target samples, the model is encouraged to move away from parameters that support correct predictions on the data to be forgotten. The objective is defined as
\begin{equation}
\mathcal{L}_{\text{GA}} = -\mathcal{L}(\mathcal{D}_f, \theta),
\end{equation}
where $\mathcal{L}(\mathcal{D}_f, \theta)$ denotes the cross-entropy loss computed on the forget set. By optimizing this objective, the model reduces its ability to answer questions related to the target private information. However, since GA does not explicitly constrain the model's behavior on non-forget data, it may also degrade overall utility and harm the retention of related public information.

\subsubsection{Negative Preference Optimization}
Negative Preference Optimization (NPO) \cite{zhangnegative} formulates unlearning as a preference optimization problem by discouraging the model from assigning high probability to target responses on the forget set $\mathcal{D}_f$. Compared with directly maximizing the forget loss, NPO provides a smoother optimization objective by contrasting the current model with a reference model. The objective is defined as
\begin{equation}
\small
\mathcal{L}_{\text{NPO}}
=
\mathbb{E}_{(x,y)\in \mathcal{D}_f}
\Bigl[
\log\!\Bigl(
1 +
\Bigl(
\frac{\pi_{\theta}(y|x)}
{\pi_{\mathrm{ref}}(y|x)}
\Bigr)
\Bigr)
\Bigr]
\end{equation}
where $\pi_{\theta}(y|x)$ denotes the conditional probability assigned by the current model, $\pi_{\mathrm{ref}}(y|x)$ denotes that of the reference model. By optimizing this objective, the model is encouraged to reduce its preference for the target outputs on the forget set while maintaining more stable updates than standard gradient ascent.

\subsubsection{KL Minimization}
KL Minimization \cite{liu2025protecting,nguyen2020variational} combines forgetting on the forget set $\mathcal{D}_f$ with a distribution-matching regularization term on the retain set $\mathcal{D}_r$. Specifically, it maximizes the loss on the forget set while encouraging the current model to remain close to the original model on retain samples by minimizing the Kullback--Leibler (KL) divergence between their output distributions. The objective is defined as
\begin{equation}
\small
\mathcal{L}_{\text{KL}}
=
-\mathcal{L}(\mathcal{D}_f, w)
+
\frac{1}{|\mathcal{D}_r|}
\sum_{s \in \mathcal{D}_r}
\mathrm{KL}\!\left(M_o \,\|\, M_c\right)(s),
\end{equation}
where $\mathcal{L}(\mathcal{D}_f, w)$ denotes the cross-entropy loss on the forget set, $M_o$ and $M_c$ denote the original model and the current model, respectively, and $\mathrm{KL}(M_o \,\|\, M_c)(s)$ measures the KL divergence between their output distributions on retain sample $s$. By optimizing this objective, the model is encouraged to forget the target private information while preserving its behavior on retain samples.

\subsubsection{Gradient Difference}
Gradient Difference (GD) \cite{liu2022continual} balances forgetting on the forget set $\mathcal{D}_f$ with the preservation of performance on the retain set $\mathcal{D}_r$. The objective is to increase the loss on $\mathcal{D}_f$ while minimizing the impact on $\mathcal{D}_r$. The overall loss function is defined as
\begin{equation}
\mathcal{L}_{\text{GD}} = -\mathcal{L}(\mathcal{D}_f, \theta) +  \lambda_{r}\mathcal{L}(\mathcal{D}_r, \theta),
\end{equation}
where $\mathcal{L}(\mathcal{D}_f, \theta)$ and $\mathcal{L}(\mathcal{D}_r, \theta)$ denote the cross-entropy loss computed on the forget set and the retain set, respectively. By optimizing this objective, the model selectively forgets the target data while preserving general utility on unrelated samples.

\subsection{Question-Answer Templates}
\label{sec:QATemplate}
For each private individual, we have eight attributes: name, salary, height, residence, employment, education, birthplace, and birthday. For each attribute, we construct a template pool containing 10 question-answer format pairs. During dataset construction, we sample from these templates to generate QA pairs. Each template includes a position placeholder, which specifies whether the target individual (or the celebrity) appears on the left or right side of the image, and an answer placeholder corresponding to the ground truth. Some examples of the question templates are shown below.

\begin{tcolorbox}[
    title=Question Template Examples,
    colback=gray!5,
    colframe=gray!40!black,
    fonttitle=\bfseries
]
\textbf{Name.}

\texttt{Question: What is the name of the person on the <position> side of the image?}

\texttt{Ground Truth: The individual on the <position> side of the image is named <answer>.}

\vspace{0.5em}
\textbf{Birthday.}

\texttt{Question: On what date was the person on the <position> side of the image born?}

\texttt{Ground Truth: The individual on the <position> side was born on <answer>.}

\vspace{0.5em}
\textbf{Height.}

\texttt{Question: What is the height of the person depicted on the <position> side of the image?}

\texttt{Ground Truth: The individual on the <position> side has a height of <answer>.}
\end{tcolorbox}

\section{Related Work Details}
\label{sec:related_work_detail}

\noindent\textbf{LLM Unlearning.} LLM unlearning aims to remove specific knowledge from a trained language model without retraining it from scratch~\cite{yao2024machine,liu2024machine,yao2024large}. Existing methods typically achieve this by training model parameters with objectives such as gradient ascent~\cite{liu2022continual}, negative preference optimization (NPO)~\cite{zhangnegative} or gradient difference \cite{liu2022continual}. To evaluate the unlearning effectiveness, several text-based unlearning benchmarks have been proposed, covering harmful knowledge~\cite{rafailov2023direct,li2024wmdp}, sensitive personal information~\cite{patil2024can,mainitofu}, and copyrighted content~\cite{eldan2023s}. For example, TOFU~\cite{mainitofu} constructs 200 fictitious author profiles with attributes such as names, birthplaces, parents' names, occupations, and written books, resulting in 4,000 question-answer pairs. WMDP~\cite{li2024wmdp} contains 3,668 multiple-choice questions for hazardous knowledge removal. These benchmarks measure whether textual LLMs can effectively unlearn the target information. However, they focus on the text modality and cannot evaluate whether unlearning methods can remove visual private information, such as facial identity in multimodal models.

\noindent\textbf{MLLM Unlearning.}
This limitation has motivated recent studies on unlearning for MLLMs, where sensitive information may exist in both text and visual content \cite{li2024single, liu2025protecting,huo2025mmunlearner,xu2025pebench,li2024single,wu2025lanp}. These studies often build upon standard LLM unlearning objectives, such as gradient ascent~\cite{liu2022continual}, negative preference optimization (NPO)~\cite{zhangnegative}, and gradient difference~\cite{liu2022continual}, while adapting them to multimodal settings that involve visual information. For example, Single Image Unlearning (SIU)~\cite{li2024single} finetunes the model on a single image for a few steps to efficiently erase visual features. MMUNLEARNER~\cite{huo2025mmunlearner} reformulates the unlearning objective to suppress visual patterns while preserving relevant textual knowledge.
Several benchmarks have been proposed to evaluate whether MLLM unlearning can effectively remove sensitive information \cite{liu2025protecting,dontsov2025clear,xu2025pebench}. MLLMU-Bench~\cite{liu2025protecting} constructs fictional personal profiles to evaluate whether unlearned MLLMs can forget sensitive information. CLEAR~\cite{dontsov2025clear} extends the TOFU benchmark~\cite{mainitofu} to the multimodal setting using images generated with PhotoMaker~\cite{li2024photomaker}. PEBench~\cite{xu2025pebench} similarly relies on synthetic profiles but extends the contexts with more scenarios, such as event scenes. Despite these advances, existing MLLM unlearning benchmarks predominantly focus on isolated or independently defined subjects and fail to capture the complex interactions between multiple subjects across multimodal contexts, leaving an important gap in evaluating the robustness of unlearning methods under realistic settings.

\subsection{Forget effectiveness of Llava Model}
We further evaluate the forget effectiveness of different unlearning methods on LLaVA-1.5-7B \cite{liu2023visual}. As shown in Table~\ref{tab:llava_forget_effectiveness}, the results are consistent with the main Qwen results: existing unlearning methods substantially reduce the model forget set performance, while public-information-preserving methods such as PIP and Blurring have relatively higher forget set accuracy.

\begin{table}[h]
\centering
\resizebox{\columnwidth}{!}{%
\begin{tabular}{lccc|ccc}
\hline
& \multicolumn{3}{c|}{\textbf{Unseen Images}} & \multicolumn{3}{c}{\textbf{Seen Images}} \\
\textbf{Method} & \textbf{Acc} (\%) & \textbf{BLEU} & \textbf{Rouge-L} & \textbf{Acc} (\%) & \textbf{BLEU} & \textbf{Rouge-L} \\
\hline
Vanilla Model       & 41.9 & 0.9022 & 0.9195 & 54.4 & 0.9169 & 0.9379 \\
Gradient Ascent     & \textbf{2.1} & 0.6832 & 0.7891 & 3.8 & 0.7034 & 0.8113 \\
NPO                 & 4.0  & 0.7026 & 0.7832 & 3.2 & 0.6867 & 0.7787 \\
KL-Min              & 5.3  & 0.6991 & 0.7870 & 4.2 & \textbf{0.6852} & \textbf{0.7735} \\
Gradient Difference & 2.3  & \textbf{0.6656} & \textbf{0.7594} & \textbf{1.7} & 0.6979 & 0.7837 \\
\hdashline
PIP (adjacent set)  & 18.1 & 0.8470 & 0.8849 & 14.2 & 0.7722 & 0.8453 \\
PIP (retain set)    & 15.2 & 0.8223 & 0.8591 & 11.5 & 0.7973 & 0.8618 \\
Blurring            & 8.6  & 0.7630 & 0.8415 & 7.6  & 0.7537 & 0.8401 \\
\hline
\end{tabular}}
\caption{Forget effectiveness of different unlearning methods on the LLaVA model under the unseen-image and seen-image test sets. Lower values indicate better unlearning performance.}
\label{tab:llava_forget_effectiveness}
\end{table}

\subsection{Hyperparameters settings}
 
Table \ref{tab:mllm_hyperparameters} shows details of hyperparameters used for different unlearning methods and MLLMs.

\begin{table*}[t]
\centering
\resizebox{\textwidth}{!}{%
\begin{tabular}{lccccccccccc}
\toprule
\multirow{2}{*}{\textbf{MLLM}} 
& \multirow{2}{*}{\textbf{Epoch}} 
& \multirow{2}{*}{\textbf{Batch Size}} 
& \multirow{2}{*}{\textbf{Optimizer}} 
& \multirow{2}{*}{\textbf{LoRA}} 
& \multirow{2}{*}{\textbf{Vanilla Learning Rate}} 
& \multicolumn{6}{c}{\textbf{Unlearning Model Learning Rate}} \\
\cmidrule(lr){7-12}
& & & & & 
& \textbf{GA} 
& \textbf{NPO} 
& \textbf{KL} 
& \textbf{GD} 
& \textbf{PIP}
& \textbf{Blurring} \\
\midrule
Qwen3-VL-4B-Instruct & 2 & 4 & AdamW & True & 1e-4 & 1e-5 & 1e-4 & 1e-5 & 1e-4 & 2e-5 & 1e-5 \\
LLaVA-1.5-7B         & 2 & 4 & AdamW & True & 2e-4 & 1e-5 & 1e-5 & 1e-5 & 1e-5 & 2e-5 & 1e-5 \\
\bottomrule
\end{tabular}%
}
\caption{Hyperparameters for different unlearning methods across MLLMs.}
\label{tab:mllm_hyperparameters}
\end{table*}

\end{document}